\documentclass[preprint,showpacs,preprintnumbers,amsmath,amssymb]{revtex4-2}

\usepackage{graphicx}
\usepackage{dcolumn}
\usepackage{bm}
\usepackage{color}

\begin{document}

\newif\ifplot
\plottrue
\newcommand{\RR}[1]{[#1]}
\newcommand{\intsum}{\sum \kern -15pt \int}
\newfont{\Yfont}{cmti10 scaled 2074}
\newcommand{\Y}{\hbox{{\Yfont y}\phantom.}}
\def\O{{\cal O}}
\newcommand{\bra}[1]{\left< #1 \right| }
\newcommand{\braa}[1]{\left. \left< #1 \right| \right| }
\def\Bra#1#2{{\mbox{\vphantom{$\left< #2 \right|$}}}_{#1}
\kern -2.5pt \left< #2 \right| }
\def\Braa#1#2{{\mbox{\vphantom{$\left< #2 \right|$}}}_{#1}
\kern -2.5pt \left. \left< #2 \right| \right| }
\newcommand{\ket}[1]{\left| #1 \right> }
\newcommand{\kett}[1]{\left| \left| #1 \right> \right.}
\newcommand{\scal}[2]{\left< #1 \left| \mbox{\vphantom{$\left< #1 #2 \right|$}}
\right. #2 \right> }
\def\Scal#1#2#3{{\mbox{\vphantom{$\left<#2#3\right|$}}}_{#1}
{\left< #2 \left| \mbox{\vphantom{$\left<#2#3\right|$}}
\right. #3 \right> }}

\title{Comprehensive investigation of the symmetric space-star configuration
  in the nucleon-deuteron breakup}

\author{H.\ Wita{\l}a$^{1}$}
\email{henryk.witala@uj.edu.pl}
\author{J.\ Golak$^{1}$}
\author{R.\ Skibi\'nski$^{1}$}
\author{K.\ Topolnicki$^{1}$}
\author{E.\ Epelbaum$^{2}$}
\author{H.\ Krebs$^{2}$}
\author{P.\ Reinert$^{2}$}
\affiliation{
$^{1}$M. Smoluchowski Institute of Physics, Jagiellonian University,
PL-30348 Krak\'ow, Poland}
\affiliation{
$^{2}$Institut f\"ur Theoretische Physik II, Ruhr-Universit\"at
  Bochum, D-44780 Bochum, Germany}

\date{\today}

\begin{abstract}
  We examine a description of  available cross section data for
  symmetric space star (SST) configurations
  in the neutron-deuteron (nd) and proton-deuteron (pd) breakup reaction using
  numerically exact solutions of the three-nucleon (3N) Faddeev equation
  based  on two- and three-nucleon (semi)phenomenological and chiral forces.
  The predicted SST cross sections are very stable with respect to the
   underlying
  dynamics for incoming nucleon laboratory energies below $\approx 25$~MeV.
  We discuss possible origins of the surprising discrepancies between theory
  and data found in low-energy nd and pd SST breakup measurements.
\end{abstract}

\maketitle

\section{Introduction}

Since the advent of numerically exact 3N continuum Faddeev 
calculations the elastic nucleon-deuteron (Nd)
 scattering and deuteron breakup reactions  
 have become a powerful tool to test  modern models of
  nuclear forces \cite{glo96,hanover,pisa}. 
  With the appearance of high precision (semi)phenomenological
  nucleon-nucleon (NN)
 potentials and first models of the 3N force (3NF) the question
 about the importance of the 3NF has become the main topic of studies in the  
  3N system. That issue was given a new impetus 
 from chiral perturbation theory (ChPT), which opened the possibility 
 to employ consistent two- and many-body nuclear forces derived within 
 this framework in 3N continuum calculations.

First applications of (semi)phenomenological  NN and 3N forces 
 to elastic Nd scattering and to the nucleon-induced deuteron 
 breakup reactions revealed interesting cases of discrepancies between
 theoretical predictions based solely on two-nucleon (2N) potentials and data, 
 indicating a possibility of large 3NF
 effects \cite{wit2001,kuros2002}. The exclusive breakup reaction offers a rich
 spectrum of kinematically complete geometries and  
the SST configuration from the very
beginning attracted attention as a possible candidate
to reveal significant 3NF effects. In this kinematically complete geometry
of  Nd breakup the momenta of the three outgoing nucleons have the same 
magnitudes and they form a three-pointed "Mercedes-Benz" 
star perpendicular to the beam direction in the 3N center-of-mass (c.m.)
 frame. 
Measurements of the pd  and nd
breakup performed at low incoming nucleon energies in different laboratories
 indeed revealed large
 discrepancies between  predicted theoretical cross sections
 and data for that geometry.
 The SST measurements 
 for nd breakup 
 have been performed at the
following energies: $E=10.25$~MeV Bochum~\cite{stephan1989},
$E=10.5$~MeV Erlangen~\cite{gebhard1993nphys} and TUNL~\cite{macri2004},
$E=13.0$~MeV Erlangen~\cite{strate1988,strate1988nphys} and
TUNL~\cite{setze1996plb,setze2005,macri2004},
$E=16.0$~MeV TUNL~\cite{crowell2001,couture2012}, 
$E=19.0$~MeV TUNL~\cite{couture2012}, and  
$E=25.0$~MeV CIAE~\cite{zhou2001}.
For the corresponding pd breakup reaction, data were taken
at $E=10.5$~MeV K\"oln~\cite{grossmann1996},  
 $E=13.0$~MeV K\"oln~\cite{rauprich1991} and
 Fukuoka~\cite{ishida2003,sagara_review_2010}, 
 $E=19.0$~MeV K\"oln~\cite{patberg1996}, and $E=65.0$~MeV PSI~\cite{psi65sst}.
 The data  provided by different groups at the same energy are consistent
 with each other both for the pd and nd systems, with nd SST cross sections
 clearly different from and larger than the pd ones.

 First analyses of the data performed in the framework of the
  3N Faddeev formalism
 with the Bonn and Paris potentials \cite{paris,bonn} showed that theory
 underpredicted
 the low-energy SST nd cross sections while simultaneously overpredicting
 the pd ones. 
 The development of NN potentials resulted in the construction of
 high precision (semi)phenomenological interactions, such as Av18~\cite{AV18},
 CDBonn~\cite{cdb}, Nijm1, and Nijm2 ~\cite{nijm}, and their
 application in 3N continuum calculations confirmed the SST
 discrepancies between theory and
 data found with the older potentials. Further progress in numerical techniques
 permitted to include the 3NF in 3N Faddeev calculations~\cite{hub97} and 
 first reported results with standard 3NF models such as Urbana IX \cite{uIX}
 or 2$\pi$-exchange Tucson-Melbourne (TM99) \cite{TM99} exhibited only
 moderate 3NF effects \cite{wit2001,kuros2002} at low energies.

 All analyses of the pd  breakup performed by the Cracow-Bochum group have 
 a permanent drawback: they neglected
 the proton-proton (pp) long-range Coulomb interaction
 present in the pd system and were based on nd calculations.  
 The successful implementation of the long-range proton-proton (pp)
 Coulomb force
 in the Faddeev formalism achieved in Ref.~\cite{deltuva2005} permitted, for the
 first time, to perform exact calculations of pd breakup.
 It turned out that 
 the pp Coulomb interaction effects are practically negligible for the SST cross
 sections \cite{deltuva2005}, which validated the results of analyses of
  that geometry in pd breakup based on nd calculations.

 Recent progress in constructing nuclear forces within chiral effective field
 theory   \cite{epel_nn_n3lo,epel2006,machpr} 
 resulted in several high precision NN potentials. Also, for the first time, a
 possibility appeared of applying  2N and 3N forces 
 derived consistently within the same
 formalism. Understanding of nuclear spectra and reactions based on these 
  consistent chiral two- and many-body forces has grown into a main topic
 of present day few-nucleon and many-nucleon studies.

 In view of these new developments it is timely to 
 check if
  it is possible to get new insights about
  the origin of the discrepancies between theory and data for the SST
  configuration by using the newly developed 
  chiral two- and three-body interactions.

The paper is organized as follows: in Sec. \ref{res_phenom} we 
present the available SST data together with their description
 by (semi)phenomenological
forces and additionally discuss sensitivity of the SST cross sections 
to particular NN force components. Results with chiral NN potentials alone 
are presented in Sec. \ref{results_chir_nn}, while in
   Sec. \ref{results_chir_3nf} the importance of 
   3NF is discussed. In Sec. \ref{discussion} we surmise on possible
   origins for the low-energy discrepancies between theoretical
   predictions and SST cross section data.
   We summarize and conclude in Sec. \ref{summary}.

\section{Results with (semi)phenomenological forces}
\label{res_phenom}

Theoretical predictions which are shown in the present paper are obtained
within the  3N Faddeev formalism using various 2N and 3N
forces. The formalism itself and information regarding our 
numerical performance were presented
in numerous publications so for details we
refer the reader to ~\cite{wit88,glo96,hub97,book}.

\begin{figure}[htbp] 
\includegraphics[scale=0.6]{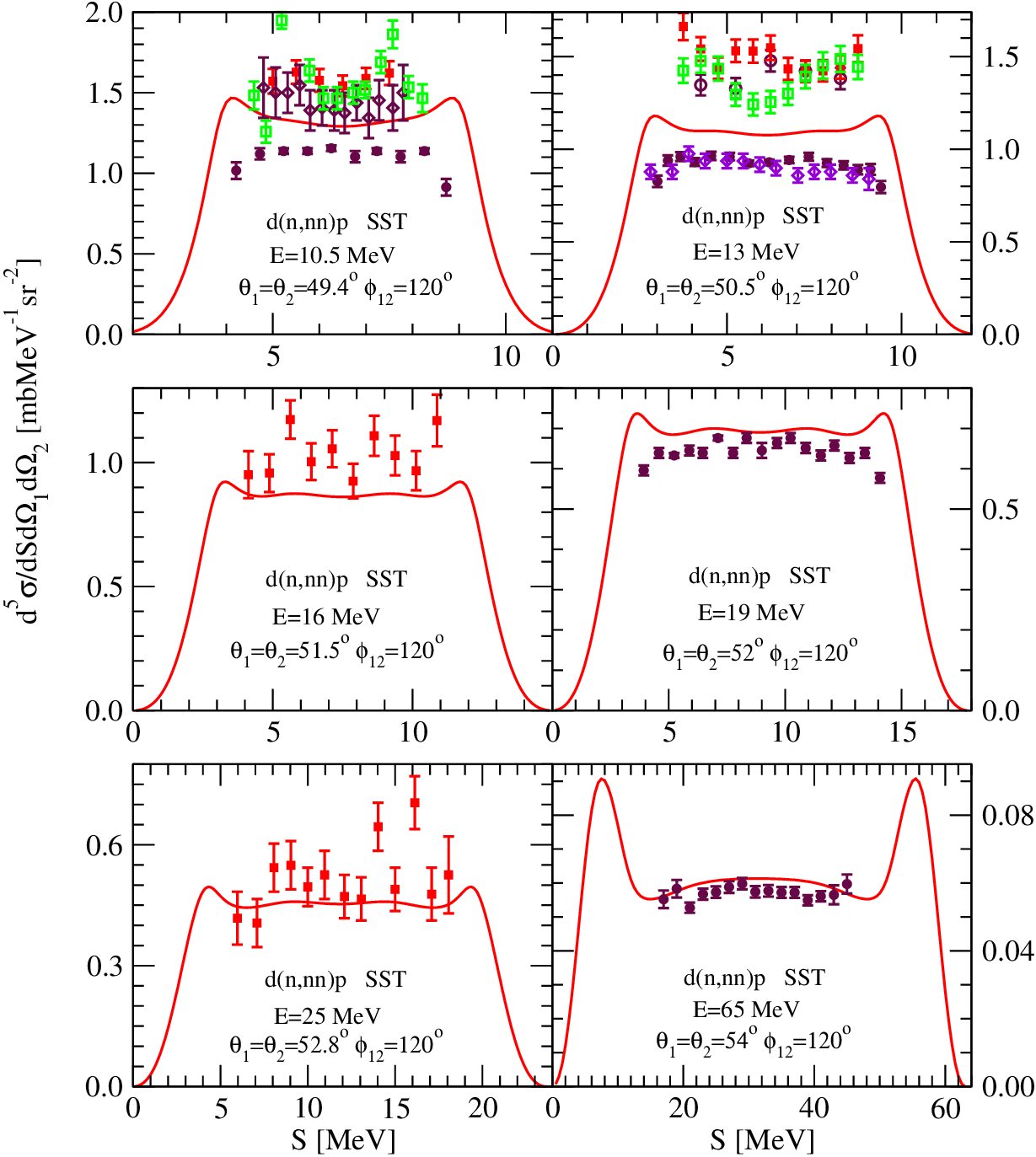}
\caption{(Color online) 
The nd breakup five-fold differential 
 cross section in the SST complete geometry at the  
 incoming neutron laboratory  energies $E=10.5, 13, 16, 19, 25$, and $65$~MeV, 
 shown as a function of arc-length of the S-curve.
 The solid (red) lines are predictions of 
 the CD~Bonn NN potential.
 At $E=10.5$~MeV the (red) squares, the (maroon) diamonds, and
 the (green) squares are 
$10.3$~MeV TUNL (\cite{macri2004}),  
$10.25$~MeV Bochum (\cite{stephan1989}), and 
 $10.5$~MeV Erlangen (\cite{gebhard1993nphys}) nd data.
 The (marron) circles are $10.5$~MeV K\"oln  (\cite{grossmann1996}) data.
 At $E=13$~MeV the (red) squares, the (maroon) circles, and
 the (green) squares are 
$13$~MeV TUNL (\cite{setze2005}),  
$13$~MeV TUNL (\cite{macri2004}), and 
 $13$~MeV Erlangen (\cite{strate1988}) nd data.
 The (maroon) circles are $13$~MeV K\"oln (\cite{rauprich1991})  
 and (violet) diamonds  $13$~MeV Fukuoka (\cite{ishida2003}) pd data. 
 At $E=16$ and $25$~MeV the (red) squares are TUNL  (\cite{crowell2001})
 and CIAE (\cite{zhou2001}) nd data.
 The (maroon) cirles at $E=19$ and $65$~MeV are  K\"oln (\cite{patberg1996})
 and PSI (\cite{psi65sst}) pd data.
 \label{fig1}}
\end{figure}

In Fig.~\ref{fig1} we show the available Nd SST cross section data and compare 
them to theory based on the CD~Bonn NN potential \cite{cdb}. It is seen
that at low laboratory energies $E$ of the incoming nucleon, 
 theoretical predictions clearly underestimate nd data by $\approx 15-30\%$
and simultaneously overestimate pd data by $\approx 10\%$. The difference
between theory and data decreases with the growing  energy, and 
at $E=65$~MeV theory describes the pd data.

\begin{figure}[htbp] 
\includegraphics[scale=0.64]{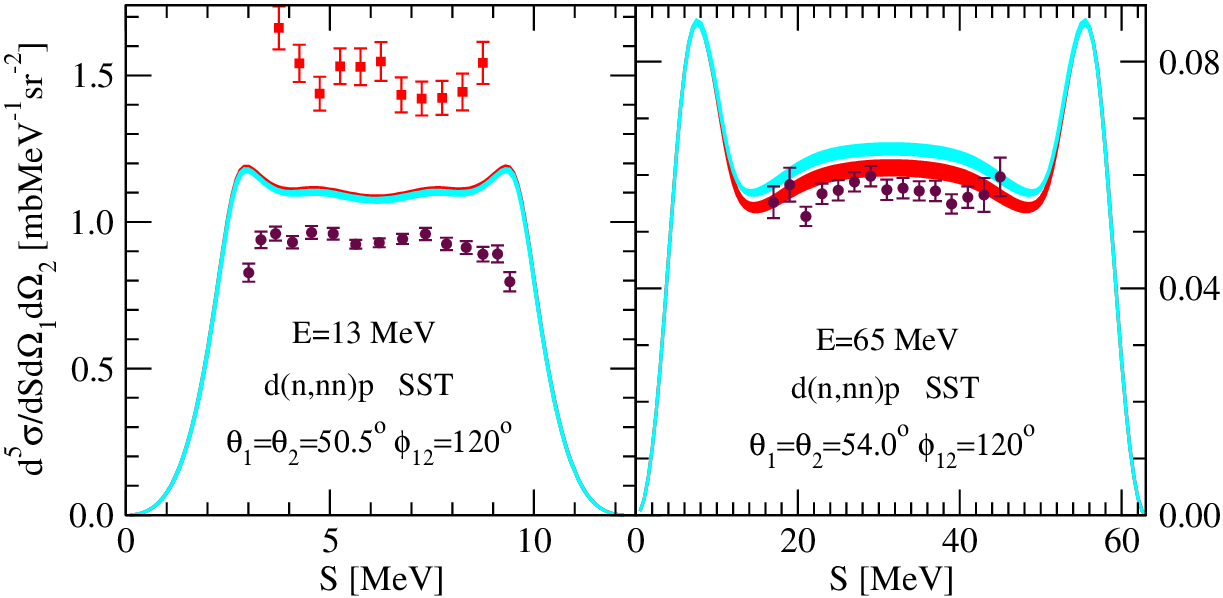}
\caption{(Color online)
  The same as in Fig.~\ref{fig1} for the incoming neutron laboratory
  energies $E=13$ and $65$~MeV. 
 The (red) dark shaded and (cyan) light shaded bands comprise predictions of 
 the AV18, CD~Bonn, Nijm1 and Nijm2 NN potentials alone or combined with
 TM99 and for AV18 also with Urbana IX 3NF, respectively.
 At $E=13$~MeV the (red) squares are 
 TUNL (\cite{setze2005}) nd data and the (maroon) circles   
 are  K\"oln (\cite{patberg1996}) pd data. 
 The (maroon) circles at  $E=65$~MeV   
 are  PSI (\cite{psi65sst}) pd data. 
\label{fig2}}
\end{figure}

The predicted  low-energy SST cross sections do not change when
instead of the CD~Bonn potential
another (semi)phenomenological interaction is used. The very narrow (red) dark
shaded band
at $13$~MeV in
Fig.~\ref{fig2}, which comprises AV18 \cite{AV18}, CD~Bonn \cite{cdb},
Nijm1 and Nijm2 \cite{nijm} predictions, reflects the astonishing
 stability of the low-energy SST cross section to the underlying dynamics.
This stability is lost at $65$~MeV as evidenced by broadening of the
band showing the predictions with NN interactions only.

The predicted low-energy SST cross sections are not only
stable with respect to the underlying NN
(semi)phe\-no\-me\-no\-lo\-gi\-cal potentials. They are also insensitive
to the standard $2\pi$-exchange 3NF's. In Fig.~\ref{fig2} we show
also (cyan) light shaded bands containing predictions based on the 
AV18, CD~Bonn, Nijm1 and Nijm2
interactions  combined with the $2\pi$-exchange
Tucson-Melbourne (TM99) 3NF \cite{TM99}. The  
 cut-off parameter $\Lambda$ of that 3NF is adjusted for each particular
NN potential and TM99 3NF combination to reproduce the experimental 
triton binding energy \cite{wit2001}. In Fig.~\ref{fig2} the band of NN+3NF
predictions contains also 
the cross section for the combination of the AV18 potential and the 
Urbana IX 3NF \cite{uIX}. 
At $13$~MeV,  effects of these 3NF's are practically negligible, and the
resulting band of predictions is very narrow and overlaps with the band
representing NN-only predictions.
The astonishing stability with respect to the
underlying dynamics present at low energies is
lost at $65$~MeV. Here, both bands  broaden significantly
 and  slightly move apart, 
indicating small effects of the 3NF. It is a region of energy where also
in elastic Nd scattering 3NF effects start coming 
into play \cite{wit2001,wit98}.

\begin{figure}[htbp] 
\includegraphics[scale=0.7]{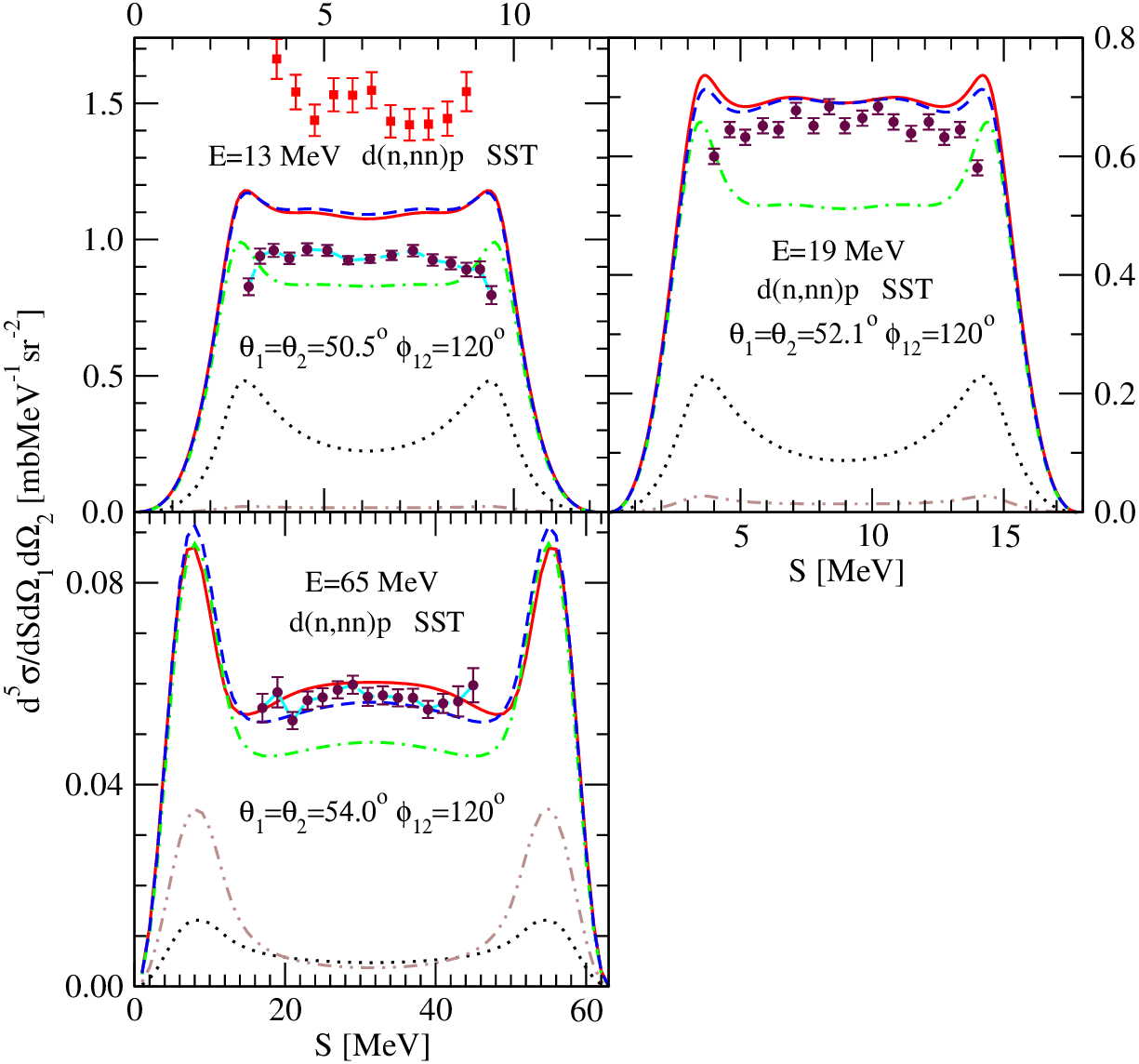}
\caption{(Color online) 
Contributions to the SST cross section at $E=13, 19$ and $65$~MeV from
  different partial waves. The (red) solid line is prediction of the
  CD~Bonn NN potential. The (black) dotted and (green) dash-dotted lines
  show cross sections obtained with 3N partial waves restricted to these which
  contain only
  $^1S_0$ and  $^3S_1$-$^3D_1$ NN partial waves, respectively.
  The  result obtained with only $^1S_0 + ^3S_1$-$^3D_1$ is given
  by (blue) dashed lines and prediction with all partial waves
  excluding $^1S_0 + ^3S_1$-$^3D_1$  is shown by  (brown)
  dash-double-dotted lines.
  The (maroon) circles at $E=19$~MeV are pd K\"oln data (\cite{patberg1996}).
  For description of the data points  at $13$ and $65$~MeV
 see Fig.~\ref{fig2}. 
\label{fig3}}
\end{figure}

The small effects of the 3NF and insensitivity to the underlying dynamics raise 
the question about the dominant NN force components, contributing to the
SST cross section. It turns out that at low energies practically 
the whole input stems from the $^1S_0$ and $^3S_1$ NN force
components only, with a dominating  $^3S_1$ contribution
 (see Fig.~\ref{fig3}). These force
components provide nearly the whole cross section at low energies as
evidenced by nearly overlapping  (red) solid (CD~Bonn prediction) and
(blue) dashed (CD~Bonn restricted to $^1S_0+^3S_1-^3D_1$ NN partial waves only)
lines at $13$ and $19$~MeV in Fig.~\ref{fig3}. 
With  increasing energy the  
contributions from the remaining partial waves become visible and
at $65$~MeV they start to outweigh the $^1S_0$ part. 

The low-energy dominance of the $^3S_1$ and $^1S_0$ contributions brings up the 
question to what extent
uncertainties of these NN force components could be responsible for the
observed discrepancies between theory and data. To answer this question
we investigated changes of the SST cross section caused by varying the
strengths of the $^3S_1$ and $^1S_0$ NN force components. To this end we
multiplied the corresponding potential matrix elements by a factor $\lambda$. 
It was shown in \cite{dineu_3ncont} that changes of the $^1S_0$ interaction
induced by the $\lambda$ values in a vicinity of $\lambda =1$ do not affect
 exclusive Nd elastic scattering observables and the total
cross sections significantly. 

\begin{figure}[htbp] 
\includegraphics[scale=0.6]{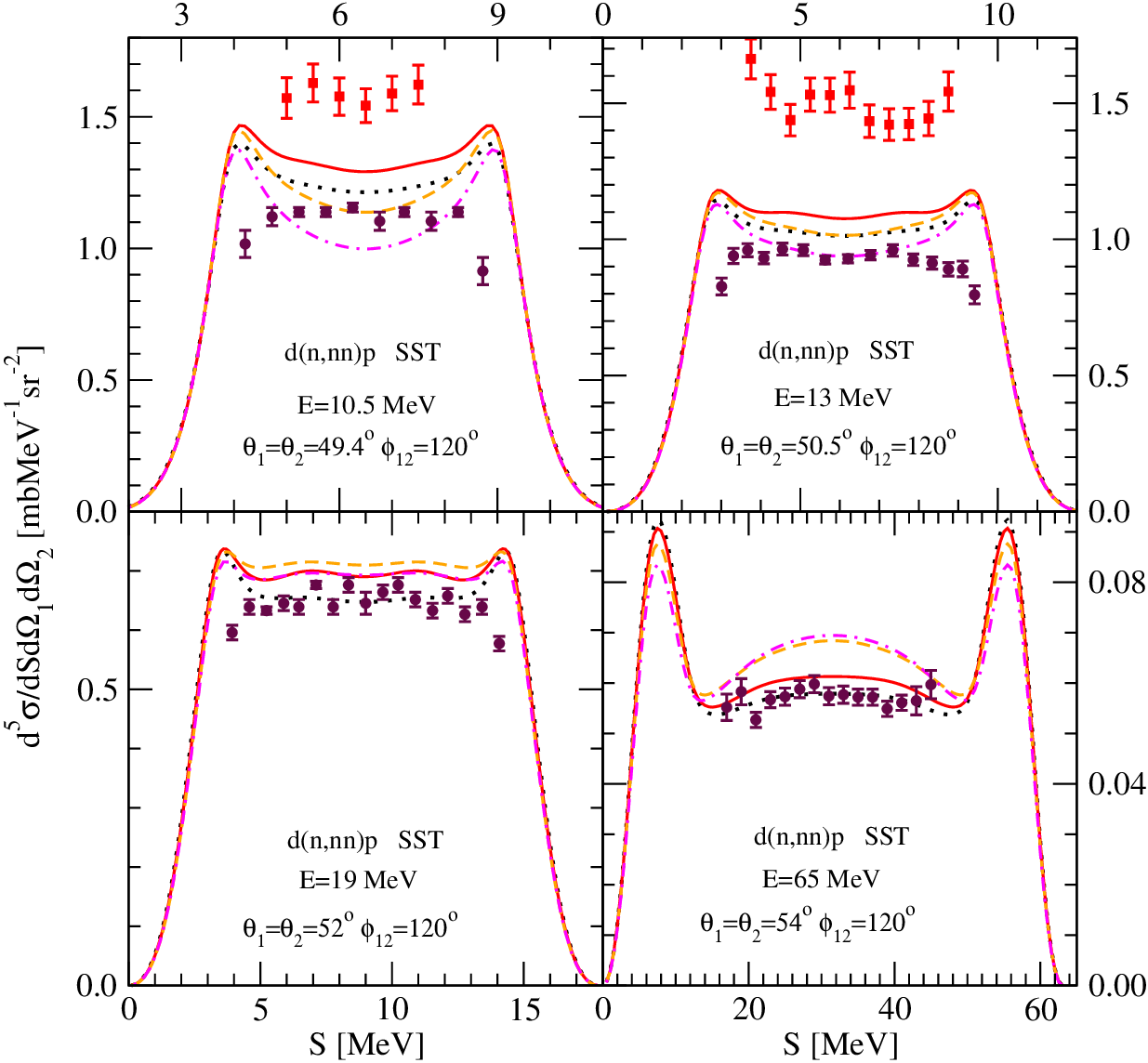}
\caption{(Color online) 
  Sensitivity of the SST cross section at $E=10.5, 13, 19,$ and $65$~MeV to
  changes of the $^1S_0$ nn force component. The (red) solid line is
  prediction of the CD~Bonn potential. The (black) dotted  curve is the
  corresponding cross section when the strength of the nn $^1S_0$ force
  component is reduced by multiplying its matrix elements
  with a factor $\lambda=0.9$.
  The (orange) dashed and (magenta) dash-dotted lines shows the cross section 
  when that strength is increased by multiplying with factor $\lambda=1.21$
  and $\lambda=1.3$, respectively. 
  For description of the data points see Fig.~\ref{fig1}.
\label{fig4}}
\end{figure}

In the first step, we investigated what changes of the  $^1S_0$ nn or pp
interaction are required to get a proper description of the low-energy
 SST cross section
data. It turned out
that it was not possible to modify the $^1S_0$ nn or pp potential in a way
which would shift the cross section predictions to the nd SST data.
However,  increasing the strength of  $^1S_0$
nn interaction by $\approx 20-30 \%$ brings theoretical predictions  close to
the pd SST cross sections at low energies (see Fig.~\ref{fig4}).
 Such a large increase of the strength ($\lambda=1.21$ or $\lambda=1.3$)
 allows two neutrons to form a $^1S_0$ bound state with the binding energy
 $E_b^{\lambda=1.21} = -0.144$~MeV or $E_b^{\lambda=1.3} = -0.441$~MeV.
Existence of such a di-neutron state does not  spoil the description of
nd elastic scattering data \cite{dineu_3ncont} but it would
 have severe consequences
 for the $^3$H binding energy, increasing it from the CD~Bonn value
 $E_{^3H}^{CD~Bonn}=-7.923$~MeV  to $E_{^3H}^{\lambda=1.21}=-9.717$~MeV or
$E_{^3H}^{\lambda=1.3}=-10.560$~MeV.
Admittedly, one could argue that an action of repulsive 3NF's 
could provide again proper binding of that system.
 However,  such a strong $^1S_0$ force  would also 
spoil the description of nuclear structure. 
 As can be seen in Fig.~\ref{fig4}, such modifications of  the $^1S_0$ nn force
 component would also lead to a significant overestimation of the pd SST cross
 section data at $E=65$~MeV but this could change upon including the 3NF tuned
 to the $^3$H binding energy.

\begin{figure}[htbp] 
\includegraphics[scale=0.7]{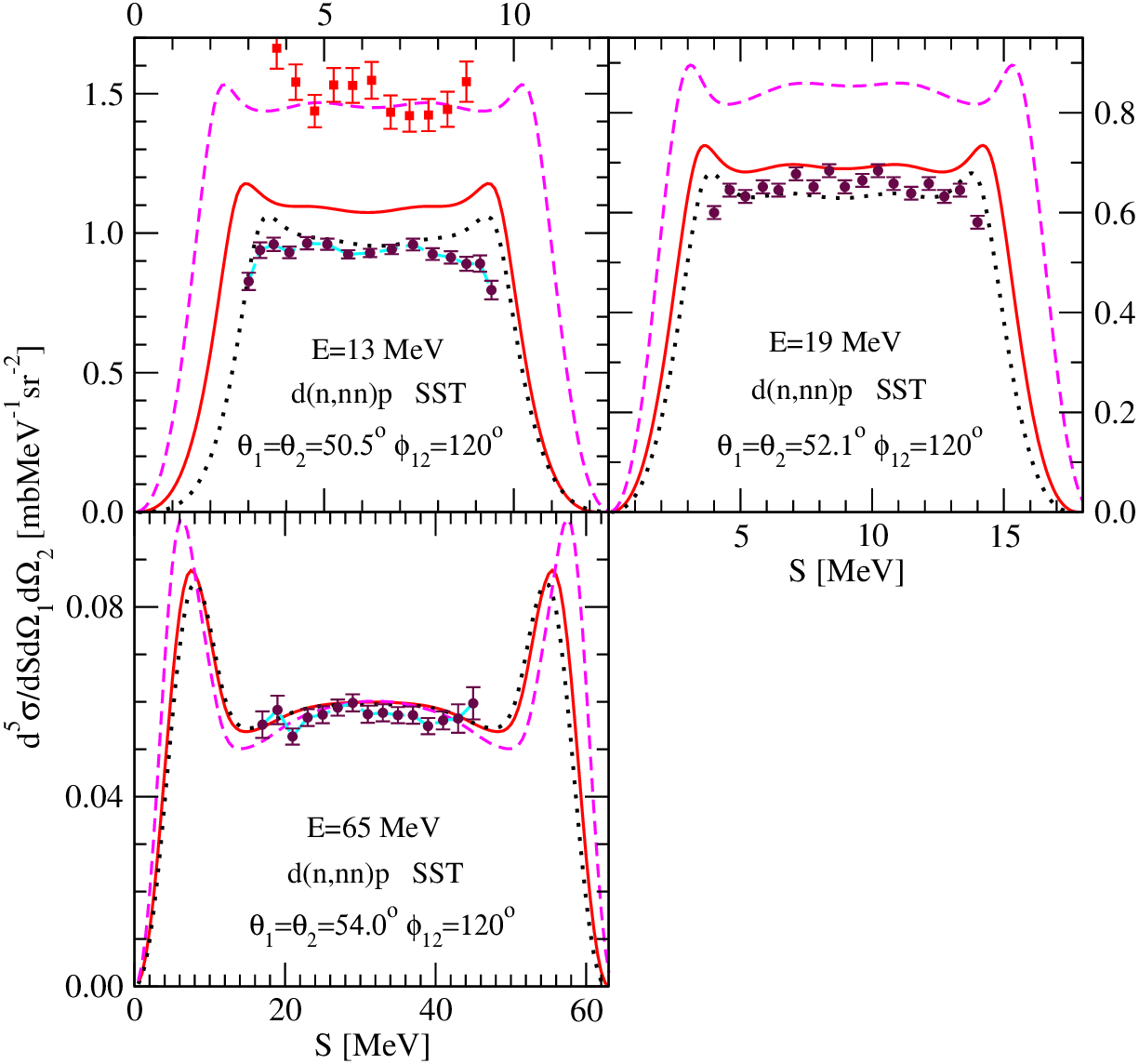}
\caption{(Color online)
  Sensitivity of the SST cross section at $E=13, 19,$ and $65$~MeV to
  changes of the $^3S_1-^3D_1$ np force component. The (red) solid line is
  prediction of the CD~Bonn potential. The (magenta) dashed curve is the
  corresponding cross section when the strength of that force component
  is reduced by $5~\%$ by multiplying its matrix elements
  with a factor $\lambda=0.95$.
  The (black) dotted line shows the cross section 
  when that strength is increased by  $2~\%$ by
  multiplying matrix elements with factor $\lambda=1.02$.
  For description of the data points see Fig.~\ref{fig4}.
\label{fig5}}
\end{figure}

 In the case of the $^3S_1$ component, a proper description
 of nd SST data  would require
a reduction  of its strength by $\approx 5 \%$ ($\lambda=0.95$)
 (see Fig~\ref{fig5}), what leads  to a complete deterioration
of the np data description and to a deuteron binding
energy $E_d^{\lambda=0.95}=-1.412$~MeV,
 drastically different from the experimental value
$E_d^{exp}=-2.224575(9)$~MeV \cite{deut_ex_bind}.
Furthermore, $^3$H would be bound by only
$E_{^3H}^{\lambda=0.95}=-6.338$~MeV and a large effect of an attractive 3NF would be
 required to regain the 
CD~Bonn $^3$H binding. On the other hand, 
the SST pd data  require only a $2\%$ increase of  $^3S_1-^3D_1$ strength
($\lambda=1.02$) (see Fig.~\ref{fig5}), what could be
still tolerated by NN data.
 However, even such a small change of the strength 
 would increase the deuteron binding to $E_d^{\lambda=1.02}=-2.592$~MeV, in
 contradiction to
 the very precise experimental value. 
 Also $^3$H would be bound stronger, with $E_{^3H}^{\lambda=1.02}=-8.598$~MeV. 

Summarizing, despite the fact that 
the SST cross sections are strongly dominated by the
S-wave NN force components, modifications of their strengths cannot 
serve to explain differences between theory and low-energy data for
that configuration. Namely those NN force components are very much restricted by
available 2N and 3N data and their variations have to be considered
with great caution.

\section{Results with chiral NN potentials}
\label{results_chir_nn}

From the available chiral NN interactions we choose four of the
 most advanced potentials
 which provide a satisfactory description of NN data in a large energy range.  
One is an older set of Bochum forces \cite{epel_nn_n3lo,epel2006} developed up
to fourth order (N$^3$LO) of chiral expansion. It  reproduces experimental  NN
phase-shifts  in  a  wide  energy range
 with an almost comparable 
accuracy as the high precision (semi)phenomenological NN potentials.
We employ five versions of that N$^3$LO chiral NN potential corresponding
to different sets of cut-off parameters used to  regularize
the  Lippmann-Schwinger  equation  and  in spectral function regularization,
namely $(450,500)$~MeV, $(450,700)$~MeV, $(550,600)$~MeV, $(600,500)$~MeV,
and $(600,700)$~MeV, denoted in the following by 201, 202, 203, 204,
and 205, respectively.

Two other choices are the new-generation chiral NN potentials
introduced and developed up to N$^4$LO  by the
Bochum-Bonn \cite{new1,epel2} and Idaho-Salamanca \cite{entem2017} groups.
While in the Idaho-Salamanca force, the nonlocal momentum-space regularization
was applied with a cutoff parameter $\Lambda$, in the Bochum-Bonn 
potential the one-pion  and two-pion exchange contributions 
are regularized in coordinate space using the cutoff parameter $R$, 
and for the contact interactions a simple Gaussian nonlocal 
 momentum-space regulator with the cutoff $\Lambda = 2R^{-1}$ was used.
 The Idaho-Salamanca N$^4$LO force is available for three values of the
 cutoff parameter $\Lambda=450, 500$, and $550$~MeV, while the
 semilocal coordinate-space regularized (SCS) chiral 
Bochum-Bonn potential has been developed for five cutoff parameters 
$R= 0.8, 0.9, 1.0, 1.1,$ and $1.2$~fm. 
 Both versions provide a very
good description of the NN data set (Idaho-Salamanca) or the phase shifts and
mixing angles of the Nijmegen partial wave analysis \cite{nijmpwa}
(Bochum-Bonn), 
used to fix the low-energy constants (LECs) accompanying the NN contact
interactions. 

The last  chiral potential considered here
 is the so-called semilocal momentum-space
regularized (SMS) chiral potential of the Bochum group \cite{preinert},
 developed up to fifth order (N$^4$LO) of chiral expansion 
 and augmented by an additional (the so-called N$^4$LO$^+$) version 
 including some sixths-order terms (also the N$^4$LO Idaho-Salamanca potential
 is augmented by the same Q$^6$ contact terms). 
 In this 2N chiral force, a new momentum-space regularization scheme
 for the long-range contributions is employed, and a nonlocal
 Gaussian regulator for the minimal set of independent contact interactions
 is introduced. These new features have also been
 applied to the corresponding 3N forces at the N$^2$LO level \cite{maris2021}. 
 This new family of semilocal chiral 2N potentials provides an outstanding
 description of the NN data and is available up to N$^4$LO$^+$
 for five values of the cutoff $\Lambda = 350, 400, 450, 500$, and $550$~MeV.

\begin{figure}[htbp] 
\includegraphics[scale=0.7]{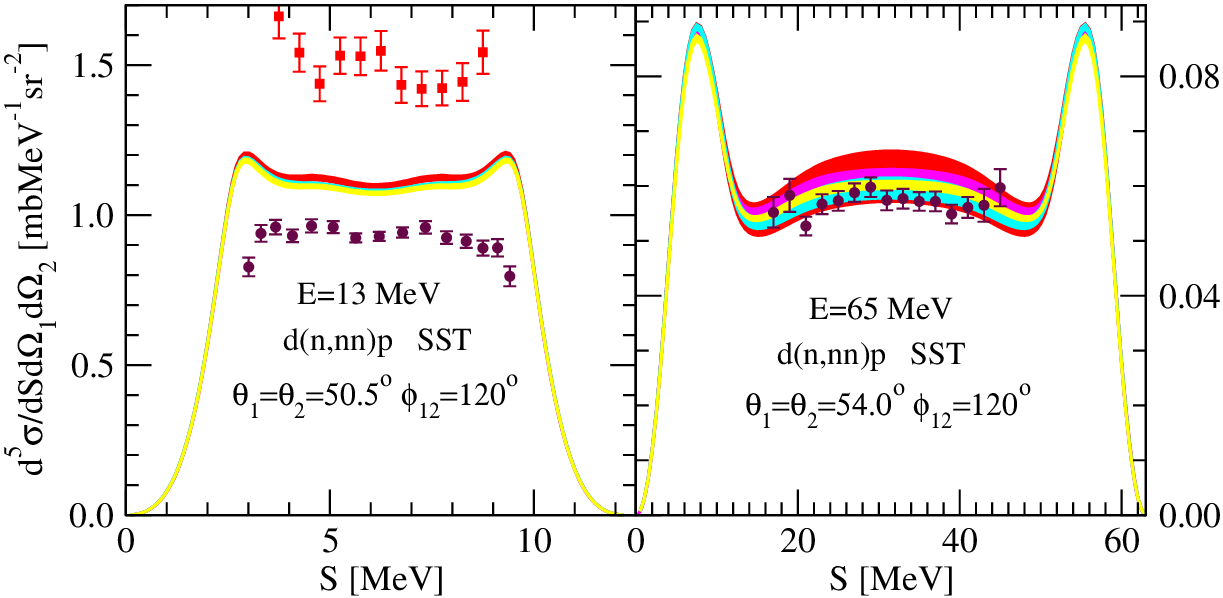}
\caption{(Color online) 
The predicted SST cross sections by  chiral NN potentials at 
incoming neutron laboratory  energies $E=13$ and $65$~MeV.
The (red) shaded band comprises five predictions of N$^3$LO chiral potentials
(201, 202, 203, 204, 205) of Ref.~\cite{epel2006} and the (magenta) band
 three predictions of N$^4$LO potentials of Ref.~\cite{entem2017}
 with regulator parameters $\Lambda=450, 500, 550$~MeV. 
The (cyan) band covers five predictions of the SCS N$^4$LO potential of
Ref.~\cite{new1} with regulator parameters $R=0.8, 0.9, 1.0, 1.1, 1.2$~fm
 and the (yellow) band four predictions of the SMS N$^4$LO$^+$
 potential of Ref.~\cite{preinert} with
 regulators $\Lambda=400, 450, 500, 550$~MeV. 
  For description of the data points see Fig.~\ref{fig2}.
\label{fig6}}
\end{figure}

 In Fig.~\ref{fig6} we show predictions of these NN potentials at $E=13$ and
 $65$~MeV  as bands comprising the available range of cutoffs or
 regulator parameters for each of the four models. At $13$~MeV the bands
 are very narrow and practically overlap with each other.
 Similarly to the (semi)phenomenological NN potentials, 
  the chiral interactions also provide
 predictions for the SST cross sections at low energies  that are very stable
 with respect to the type of the underlying interaction
 and its parameters as far as
 they provide a satisfactory description of the NN data.
 The predictions of the chiral potentials agree with those of
 (semi)phenomenological forces, leading to the same disagreement with
  the low-energy  SST cross  section data.

  As can be seen  in Fig.~\ref{fig6} these bands are broadened  at $65$~MeV,
   especially for the older Bochum-Bonn potential (the (red) dark shaded 
 band in Fig.~\ref{fig6}),  reflecting  the increased dependence of 
 the predictions on potential parameters at that energy as well as the
 worse description of NN  data by the older Bochum-Bonn potential.
 The bands of SCS and SMS interactions are significantly constricted, in line
 with a good representation of the NN phase-shifts by these potentials.

 The applied chiral potentials differ not only
 in their regularization scheme.
 The older Bochum-Bonn potential leads to
 the deuteron wave function, which is quite different from the ones 
 obtained with other chiral potentials \cite{wit_jpg}.
 Despite these differences
 the calculated low-energy SST cross sections are practically the same.

\section{Results with chiral 3N-forces}
\label{results_chir_3nf}

First nonvanishing 3NF contributions appear
at N$^2$LO \cite{vankolck,epel2002} and contain, 
in addition to the $2\pi$-exchange term, two short-range contributions with the
 strength
parameters $c_D$ and $c_E$ \cite{epel_tower}. The latter two can be determined 
from the $^3$H binding energy and the Nd 
differential cross section minimum at about $E_{lab}=70$~MeV, which
is the energy
at which effects of 3NF start to appear in the Nd elastic scattering cross
section \cite{wit2001,wit98,maris2021,epel2019}.  
Specifically, first the so-called ($c_D,c_E$) correlation line is established,
which for a particular chiral NN potential combined
with a N$^2$LO 3NF yields values of ($c_D,c_E$)  reproducing
the $^3$H binding energy. Then, a fit to the experimental data for the elastic 
Nd cross section is performed and the
values of both strengths, $c_D$ and $c_E$, are uniquely determined. 

\begin{figure}[htbp] 
\includegraphics[scale=0.6]{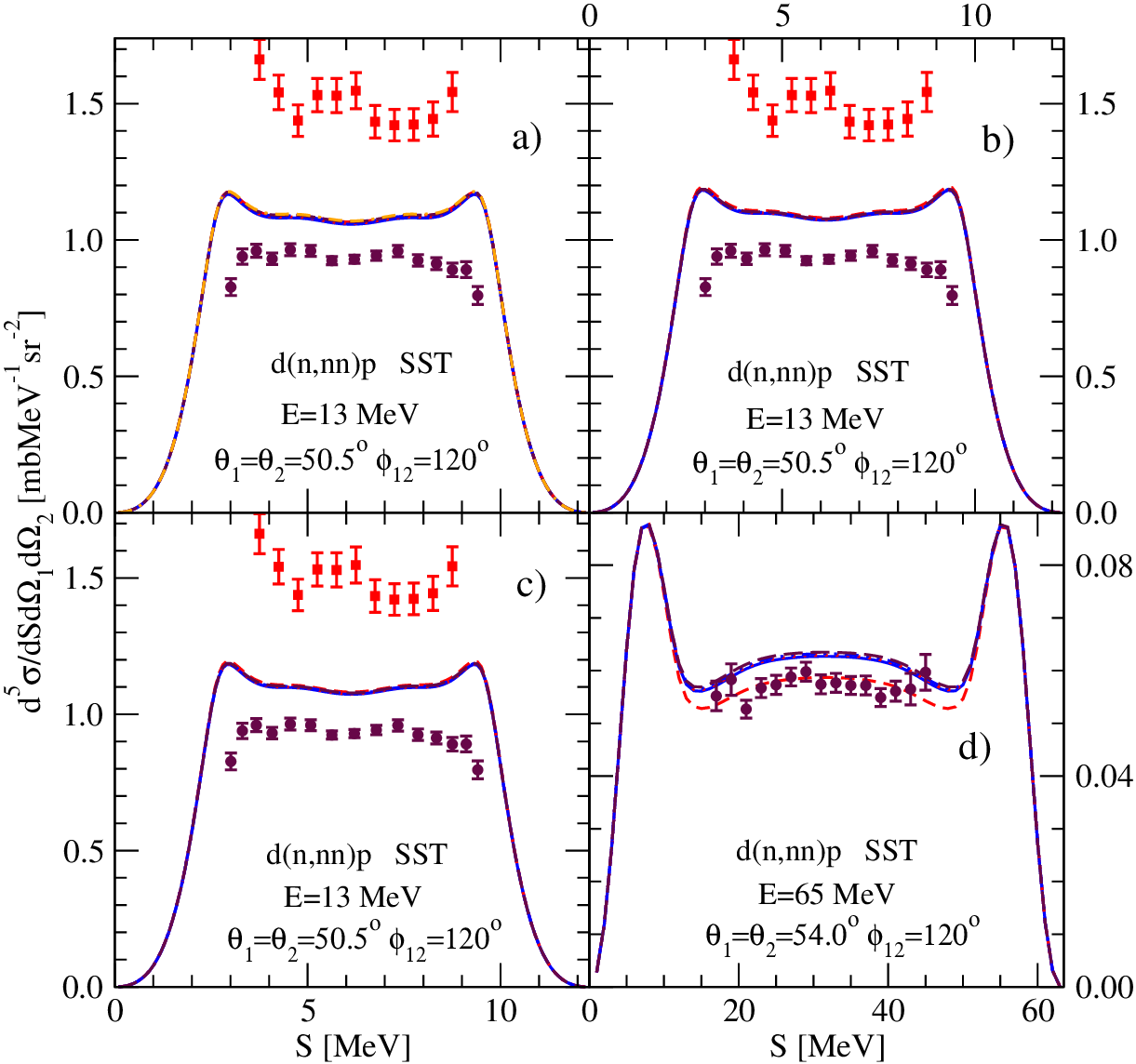}
\caption{(Color online) 
  Effects of N$^2$LO 3NF on SST cross section at  $E=13$ and $65$~MeV
  when combined with
  chiral SCS NN potential at different orders of chiral expansion. 
 The dashed (red) lines are predictions of 
 the SCS NN potential with the regulator $R=0.9$~fm at N$^2$LO (a),
 N$^3$LO (b),  and N$^4$LO (c) and (d).
 Combining that potential with the N$^2$LO 3NF with four 
 strengths  of the contact 
 terms from the correlation lines ($c_D,c_E$)  leads to results shown  
 by different curves: solid (blue)
 (a: ($-4.0,0.344$), b: ($4.0,0.103$), c,d: ($4.0,-0.270$)), 
 dotted (red) (a: ($-2.0,0.131$), b: ($6.0,-0.960$), c,d: ($6.0,-1.094$)), 
 double-dotted-dashed (blue)
 (a: ($0.0,-0.097$), b: ($8.0,-1.937$), c,d: ($8.0,-2.032$)), 
 and  dashed (maroon)
 (a: ($2.0,0.345$), b: ($10.0,-3.063$), c,d: ($10.0,-3.108$)).  
  For description of the data points see Fig.~\ref{fig2}. 
 \label{fig7}}
\end{figure}

 In Fig.~\ref{fig7}a, b, and c we show predictions for the SST cross section
 at $E=13$~MeV for the SCS chiral potential with the
regularization parameter $R=0.9$~fm at N$^2$LO, N$^3$LO, and N$^4$LO,
  respectively, combined with the N$^2$LO 3NF for four sets of
the strength parameters taken from the corresponding correlation lines. Also, 
 predictions of particular chiral potentials are shown by (red) dashed
line. All lines  practically overlap showing that effects of N$^2$LO
3NF on $13$~MeV SST cross section are negligible. The predicted cross section
is insensitive to the order of the chiral NN potential used.

Effects of the N$^2$LO 3NF start to appear at $65$~MeV
(see Fig.~\ref{fig7}d). This is the energy region, where 
effects of 3NF's start to come into play also in
elastic Nd scattering \cite{wit2001,wit98}.
The overlapping   predictions for four sets of strengths combinations
from the correlation line ($c_D,c_E$) shown in Fig.~\ref{fig7}d 
indicates, that the magnitude of 3NF effects in this energy range does
not depend on strength values as far as they 
 are taken  from the correlation line.

\begin{figure}[htbp] 
\includegraphics[scale=0.7]{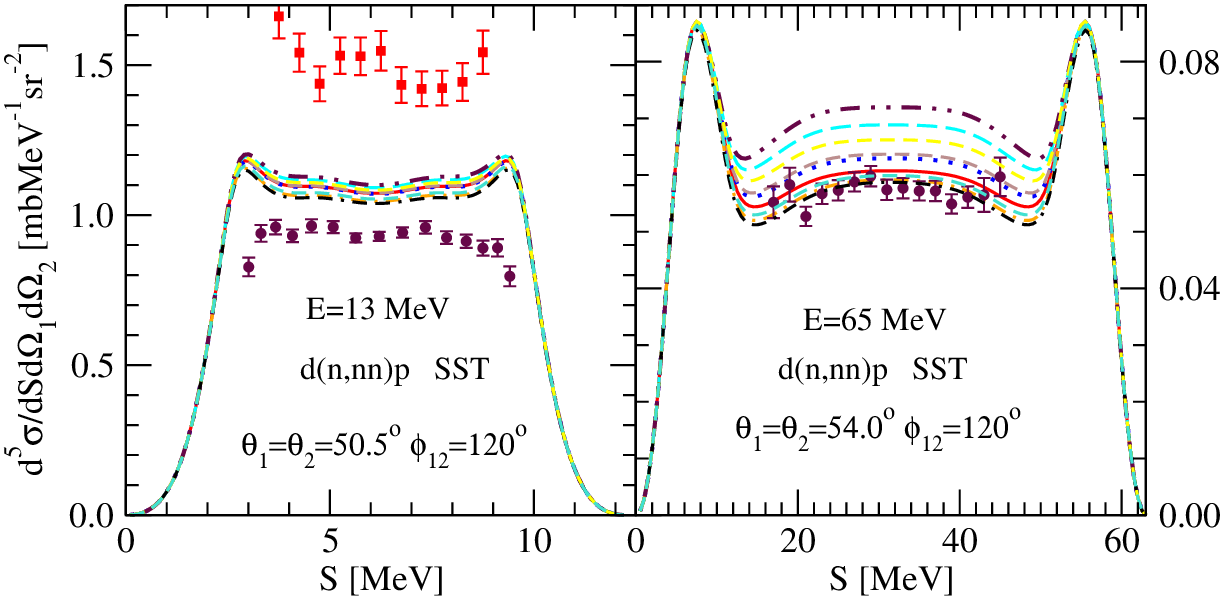}
\caption{(Color online)
  SST cross sections at  $E=13$ and $65$~MeV predicted by SMS chiral 
  N$^4$LO$^+$ NN potential (the regulator parameter $\Lambda=450$~MeV) alone
  (shown by dashed (red) lines)
   or combined with consistently regularized N$^2$LO 3NF with different
   strengths of the contact terms taken from the correlation
   line ($c_D,c_E$) and shown  
 by lines: dotted (blue) ($2.0,0.287$), 
 short dashed (brown) ($4.0,0.499$), 
 long dashed (cyan) ($15.0,1.362$), 
 dash-dotted  (orange) ($-15.0,-2.010$),
 dash-double-dotted (maroon) ($20.0,1.533$), 
 double-dash-dotted (black) ($-20.0,1.362$), 
 and short dashed  (turquoise) ($-10.0,-1.257$). 
  For description of the data points see Fig.~\ref{fig2}.
\label{fig8}}
\end{figure}

To investigate further how effects of the N$^2$LO 3NF depend on the strengths
 of the contact terms we took the most precise chiral SMS potential at 
N$^4$LO$^+$ with the regulator $\Lambda=450$~MeV and combined it with
   the N$^2$LO 3NF \cite{epel_tower}.
 In Fig.~\ref{fig8} we show predictions for SST cross sections at $E=13$
 and $65$~MeV for eight combinations of strengths taken from the
 correlation line ($c_D,c_E$). Again, in spite
 of a very wide range of $c_D$ values, taken between $c_D=-20$ and $c_D=+20$,
 the predicted $13$~MeV cross sections are found to lie within a relatively
  narrow band. Contrary to that, at
  $65$~MeV a very broad range of predictions is seen confirming
  the observation that in this energy region 3NF effects become important.

One may now raise the question of the role of 
  3NF components from higher chiral orders
and their impact on the SST cross section. 
 The necessary work to derive the chiral
 3NFs at N$^3$LO has been done in \cite{3nf_n3lo_long,3nf_n3lo_short} using
 dimensional regularization. 
 At that order, five different topologies contribute
 to the 3NF. Three of them are of long-range character \cite{3nf_n3lo_long}
 and are given by two-pion ($2\pi$) exchange graphs, by
 two-pion-one-pion ($2\pi-1\pi$) exchange graphs, and
 by the so-called ring diagrams. They are supplemented by the short-range
 one-pion-exchange-contact (1$\pi$-contact) and 
two-pion-exchange-contact (2$\pi$-contact) terms \cite{3nf_n3lo_short}. 
 The 3NF at N$^3$LO order
 does not involve any new unknown low-energy constants (LECs), see, however,
 a related discussion in \cite{girlanda2020},  and depends only
on two parameters, $c_D$ and $c_E$, that parameterize the leading
one-pion-contact term and the 3N contact term appearing already at N$^2$LO.
Their values need to be fixed at a given order 
from a fit to few-nucleon data,
as in the N$^2$LO case. 

In the first preliminary investigation of N$^3$LO 3NF
effects \cite{wit_jpg} we considered the action of the
 3NF only in 3N states with
the total 3N angular momenta $J=1/2$ and $3/2$ and included
 all long-range contributions 
 with the exception of 1/m corrections. Additionally, the
  $2\pi$-exchange-contact term was
 omitted in the short-range part of 3NF.
 The strengths parameters $c_E$ and $c_D$ were determined at that time
 from the correlation line and the nd doublet scattering length $^2a_{nd}$
 was used in addition to the $^3$H binding energy to uniquely determine both
 values. 

\begin{figure}[htbp] 
\includegraphics[scale=0.7]{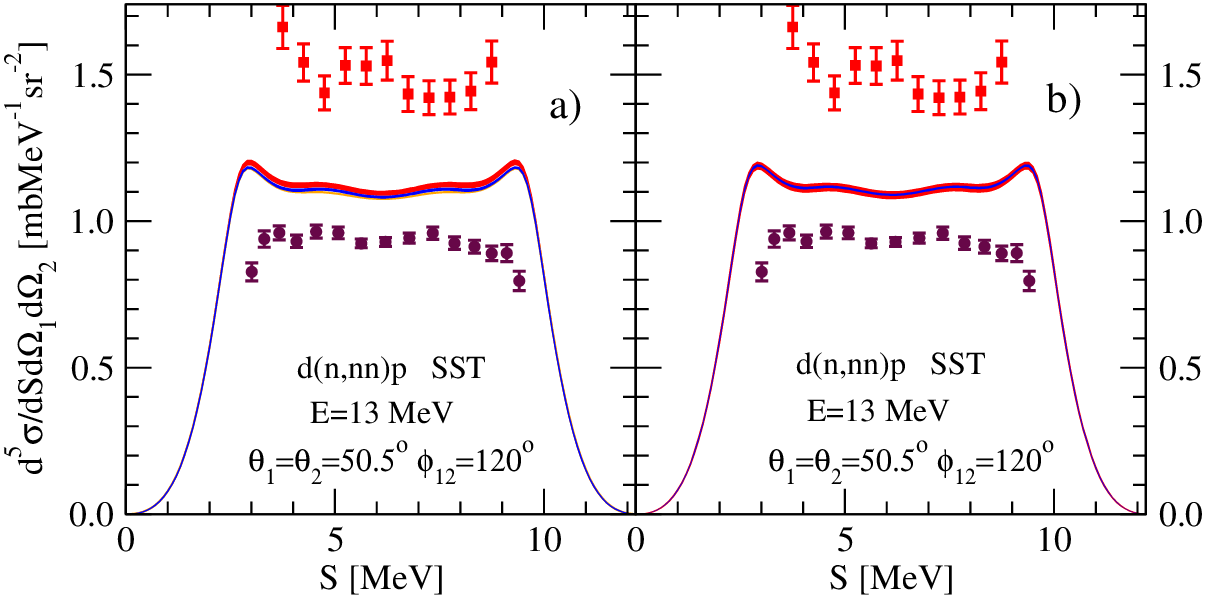}
\caption{(Color online)
  Effects of N$^3$LO 3NF, restricted to 3N total angular momenta
  $J=1/2$ and $3/2$, with all long-range contributions with
  exception of $1/m$ corrections and omitted 2$\pi$-exchange-contact
  term in the short-range part on SST cross section at  $E=13$~MeV (a). 
 The red band in (a) comprises predictions of five 
 chiral  N$^3$LO Bochum-Bonn potentials (201-205).
   Combining them with  N$^3$LO 3NF with 
 strengths   of the contact 
 terms determined by the  correlation line ($c_D,c_E$)
 and requirement to reproduce $^2a_{nd}$ leads to the blue band.
 The (orange) solid line is prediction of the CD~Bonn potential.
 In (b) the same is shown for chiral Bochum-Bonn potentials and 3NF at N$^2$LO.
 In this case the N$^2$LO 3NF acted up to  $J=7/2$.  
  For description of the data points see Fig.~\ref{fig2}.
\label{fig9}}
\end{figure}

 In Fig.~\ref{fig9}a we show the SST cross sections at $13$~MeV 
  in form of a red band comprising  five predictions of  the 
  N$^3$LO Bochum-Bonn potentials (versions 201-205).
  Combining these potentials with the N$^3$LO chiral 3NF gives the blue band. 
  For the sake of comparison also the CD~Bonn prediction is shown
  by the (orange)
  solid line. In b) the corresponding predictions at N$^2$LO are also
  presented. 
  It again turns out  that the cross section for the SST
configuration of the nd breakup is very stable with respect to
the underlying dynamics. 
Not only (semi)phenomenological potentials, alone or combined
with standard 3N forces, provide
practically the same SST cross sections. Also 
 the chiral 2N forces supplemented by the N$^3$LO 3NF
 without relativistic 1/m
 corrections and short-range $2\pi$-contact term yield similar predictions 
 and cannot explain the discrepancy
 between the theory and the data found for the low-energy SST configurations.
 Notice further that a consistent regularization of the 3NF beyond N$^2$LO has
 not yet been achieved, see \cite{epel1911}. 

Due to the restriction to the low total 3N angular momenta this result has
to be confirmed. A systematic investigation of effects of 
the 3NF beyond N$^2$LO is the main aim of the LENPIC
 collaboration \cite{epel2019}.

At fifth order (N$^4$LO) the chiral  3NF comprises thirteen  
purely short-range operators \cite{girlanda2011} in addition to the long- and
intermediate-range interactions generated by pion-exchange
diagrams \cite{krebs2012,krebs2013}. 
 In an exploratory study of \cite{girlanda2019},  effects
of these subleading short-range terms were investigated in
 pd scattering below $E_{lab} = 3$~MeV within a 
 hybrid approach based on phenomenological two- and three-nucleon forces.

\begin{figure}[htbp] 
\includegraphics[scale=0.7]{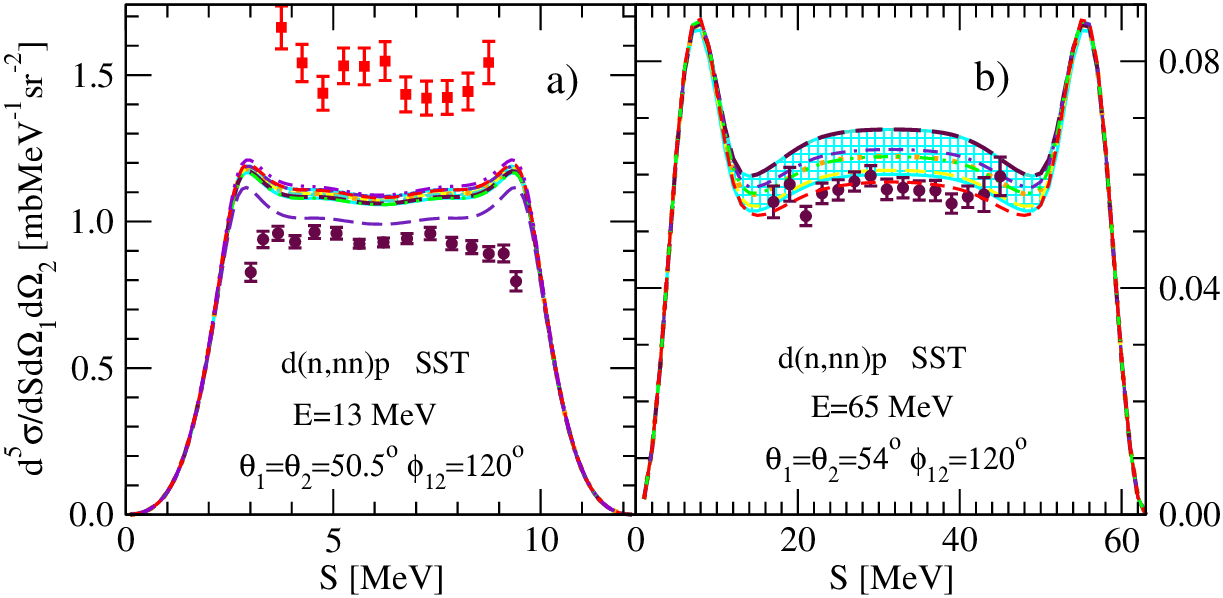}
\caption{(Color online)
SST cross sections at  $E=13$ and $65$~MeV predicted by 
 the  chiral SMS N$^4$LO$^+$ NN potential with the regulator parameter
 $\Lambda=450$~MeV alone (the long dashed (yellow) lines) or combined with
 3NF comprising N$^2$LO and two out of ten N$^4$LO contact terms,  
 with  strengths  of the contact 
 terms from the correlation line ($c_D,c_E,c_{E_1},c_{E7}$). 
 The results with 3NF are shown  
 by the following lines: dotted (orange) ($2.741,0.367,0.0,0.0$), 
 dash-double-dotted (turquoise) ($10.818,-0.295,-4.0,0.0$), 
 long dashed (maroon) ($-11.977,-0.969,4.0,0.0$), 
 double-dash-dotted  (indigo) ($10.471,1.079,0.0,-4.0$), and 
 dash-double-dotted (green) ($-8.205,-1.002,0.0,2.0$).
 The (cyan) shaded band covers region of predictions given by five above and
 three additional sets of strengths: ($6.971,0.207,-2.0,0.0$),
 ($8.533,0.925,0.0,-2.0$), ($3.133,0.041,2.0,0.0$). 
 For comparison also the short dashed (red) line is drawn which is the  
 prediction of the SCS N$^4$LO potential with the regulator parameter
 $R=0.9$~fm. In a), two additional lines are drawn which correspond to the 
 combinations of strengths ($c_D,c_E$) outside the correlation line, namely
 (indigo) long dashed: ($2.0,1.433$) and (red) dash-double-dotted:
 ($2.0,-1.433$).
  For description of the data points see Fig.~\ref{fig2}.
\label{fig10}}
\end{figure}

 To get insight into the expected 3NF effects on SST cross sections
 from these N$^4$LO short-range 3NF contributions   
   we choose two out of the ten terms, namely the
 isoscalar central and spin-orbit interactions coming with strengths $c_{E_1}$
 and $c_{E_7}$, respectively \cite{girlanda2011,epel_tower}
 and add them to the N$^2$LO 3NF. 
 In Fig.~\ref{fig10} we show predictions for the SST cross section at
 $E=13$ and $65$~MeV for the chiral SMS N$^4$LO$^+$ potential
 with regularization
 parameter $\Lambda=450$~MeV combined with that 3NF for a set of
 eight combinations of strengths from the correlation
 lines ($c_D,c_E,c_{E_1},c_{E_7}$) for fixed values of $c_{E_1}$ and $c_{E_7}$. 
  For five of these combinations also the lines of predictions are drawn.
 It is seen  that the inclusion of the $c_{E_1}$- and $c_{E_7}$-terms
  has a negligible effect on the
 $13$~MeV SST cross section  and the predicted cross sections 
  essentially coincide with each other and with the results of SCS N$^4$LO
  and SMS N$^4$LO$^+$ potentials alone.  Again, at $65$~MeV, a wide 
  spread of predictions indicates significant effects of the 3NF at this energy.
  
The observed discrepancy between  theoretical predictions and
the nd  and pd low-energy SST cross section data
thus indeed appears to be puzzling. Due to the observed strong stability of
the low-energy space-star cross sections to the
 underlying dynamics it seems very
unlikely that this puzzle can be resolved by the inclusion of omitted  N$^3$LO 
 terms or remaining N$^4$LO and higher order contributions to the 3NF.
 We checked for a combination of SMS N$^4$LO$^+$ NN and N$^2$LO 3NF that even
 removing the requirement to reproduce  $^3$H binding energy does not help
 to come into the vicinity of the $13$~MeV nd SST cross section data. Namely
 taking strengths $c_D=2.0$ and $c_E=0.287$ from the correlation line and
 increasing $c_E$ to  $c_E=1.433$ lowers the predicted cross sections and
 brings them close to pd SST data (see (indigo) long dashed
 line in Fig.~\ref{fig10}). For the combination of strengths
 $c_D=2.0, c_E=1.433$, $^3$H is strongly bound with $E_{^3H}=-10.783$~MeV. 
  However, decreasing the $c_E$ value and
 even changing its sign does not have any significant effect on the predicted
 cross  section, which remains close to the stable region of predictions
 for strength values from the correlation line (see Fig.~\ref{fig10} and
 the (red) dash-double-dotted line, which is prediction
 for $c_D=2.0$ and $c_E=-1.433$). For the combination of strengths
 $c_D=2.0, c_E=-1.433$, $^3$H is bound with $E_{^3H}=-6.792$~MeV. 

\section{Discussion of the low-energy discrepancy}
\label{discussion}

\begin{figure}[htbp] 
\includegraphics[scale=0.7]{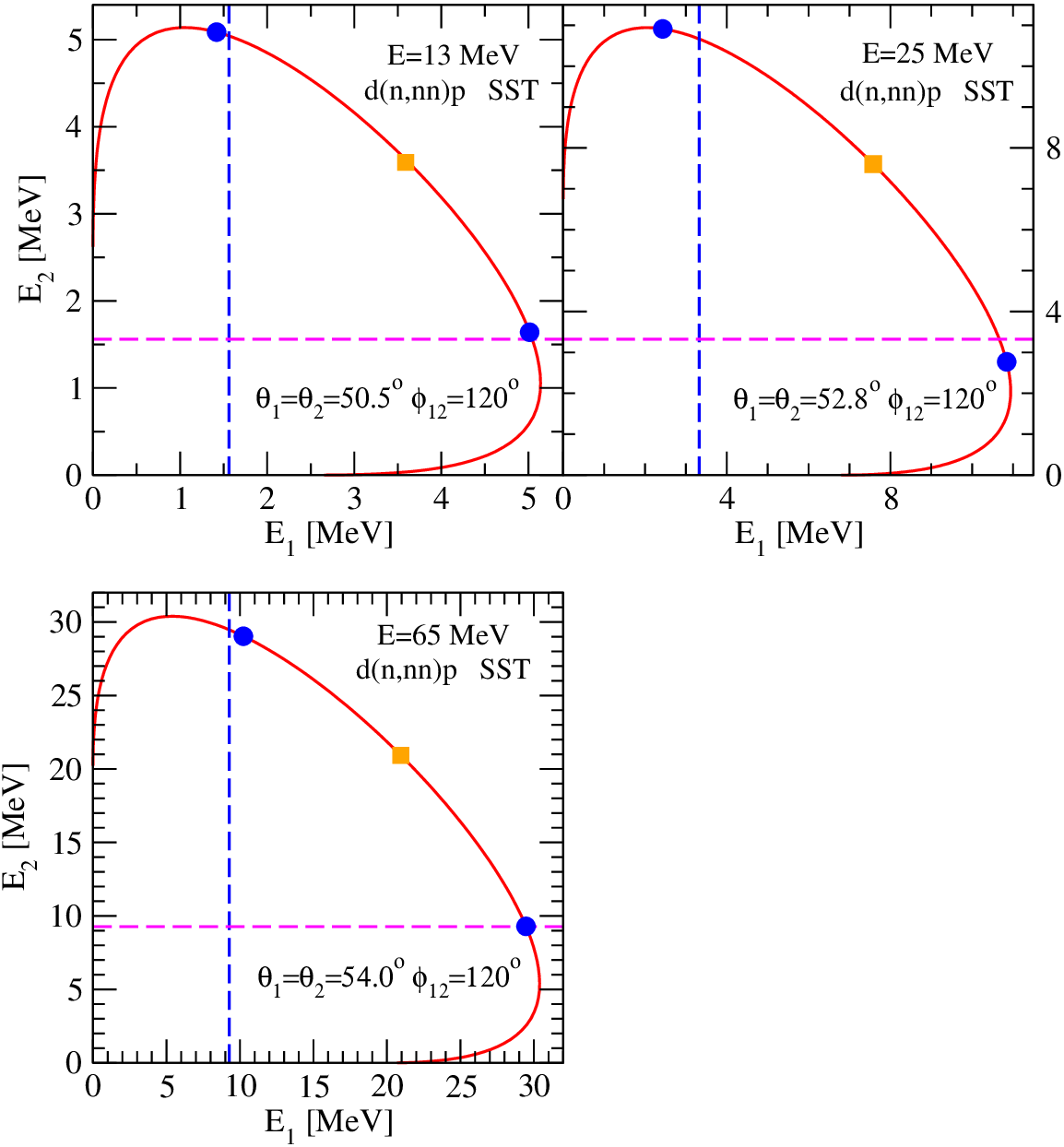}
\caption{(Color online)
  Position of $d(n,nn)p$ breakup events (S curve) in the laboratory
  kinetic energy plane $E_1-E_2$
  of two detected in coincidence neutrons for SST configurations
  at $E=13, 25,$ and $65$~MeV. The dashed lines
  show half of the energy of the outgoing di-neutron
  from $d(n,di-neutron)p$ reaction at  lab. angle of
    the corresponding SST geometry. The position on the S-curve 
  where the SST condition is fulfilled is shown by square and the range
  of S-curve covered by  nd (pd) SST data is indicated by  dots.
\label{fig11}}
\end{figure}

\begin{figure}[htbp] 
\includegraphics[scale=0.7]{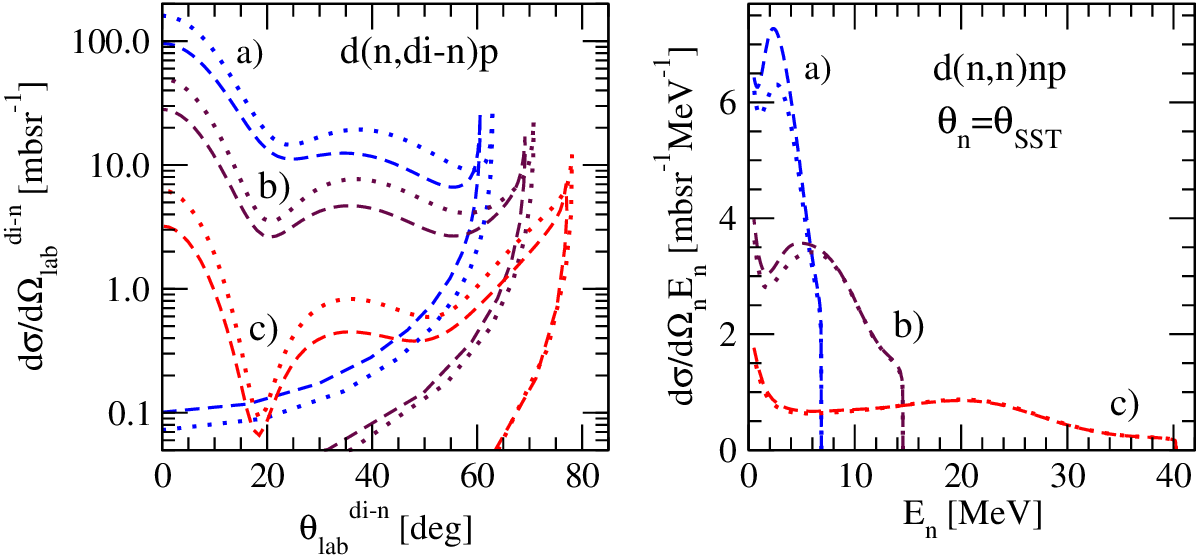}  
\caption{(Color online)
  The angular distributions of di-neutrons and 
the energy spectra of outgoing neutrons from incomplete breakup
reaction $d(n,n)np$ for lab. angle of the detected neutron equal to the
SST configuration angle,   predicted by  
the CD~Bonn potential with a $^1S_0$ nn force component modified by
increasing its strength by a factor $\lambda$ to get two neutrons bound. 
 The (blue) dashed line corresponds to $\lambda=1.21$ with
di-neutron binding energy $E_b^{di-n}=-0.144$~MeV and the dashed-dotted line
to $\lambda=1.3$ with $E_b^{di-n}=-0.441$~MeV. The a), b) and c) are
predictions for incoming neutron lab. energies $E=13, 25,$ and $65$~MeV,
respectively. 
\label{fig12}}
\end{figure}

The presented results support the conjecture that 3NF effects are not 
 responsible for the discrepancies  between data and theory in the low-energy
 SST cross sections. 
 One could argue that, perhaps, modifications of the $^1S_0$ and/or
 $^3S_1-^3D_1$ NN force components, more refined than a simple change of their
 strengths,  would provide an explanation for at least pd SST low-energy
 cross section data. However, this seems to be difficult since there is
 no room for modifications of np and pp forces compatible with NN
 data \cite{nijmpwa,navarro,preinert2006}.
 In spite of the fact that the pd SST  discrepancy is relatively
 small ($\approx 10~\%$) in comparison to the nd SST one, since the theoretical
 predictions lie well outside
 statistical error bars and  the systematic errors are claimed
 to be small \cite{sagara_review_2010}, it presents a significant discrepancy.
 A dedicated pd SST measurement aimed to determine precise normalization
 of the SST cross sections would help to put some light on this discrepancy.
 Even larger is the discreapancy to nd
 SST data and between pd and nd data themselves. 
 However, due to  the strong insensitivity
 of low-energy SST cross sections to the
 underlying dynamics it seems rather unlikely,  that any conceivable
 charge symmetry breaking mechanism in the NN and/or 3N force would be
 able to explain the difference between pd and nd SST data and allow 
 to  describe nd data. This situation poses an interesting puzzle
 for theory  and its solution has to be probably sought in 
 some exotic mechanism contributing to the nd breakup and irrelevant for the
 pd one. When looking
 for such a mechanism one could consider the contributions of
 hypothetical bound state of two neutrons in the state $^1S_0$ 
 to a region of SST breakup phase-space. Such contributions could create
 in nd SST measurements additional background originating from
 accidental coincidences
 between breakup neutrons and di-neutrons produced in nd scattering, increasing
 thus the measured cross section. That such a scenario is conceivable
 follows from the fact that detection of neutrons in nd SST
measurements was performed using liquid scintillators with pulse-shape
discrimination and their energy was determined by  the 
time-of-flight technique \cite{setze2005}.
Such a detection system does not distinguish
between di-neutrons and breakup neutrons. 
 As a result,  accidental coincidences between breakup neutrons and di-neutrons
  could appear in the region of SST
phase-space which cannot be distinguished from true events by the
applied measurement technique. The kinetic energy assigned
 to di-neutron when using such an experimental arrangement, which is 
determined by time of flight measurement of its velocity, will
 be in consequence twice as small as its real kinetic energy.
To be specific, assuming that di-neutron is bound
by $E_b^{di-n}= -0.144$~MeV (what corresponds to
the factor $\lambda_{^1S_0}=1.21$) would lead at incoming neutron lab. energy
$E=13$~MeV to the energy of outgoing di-neutrons from $d(n,di-neutron)p$
reaction at laboratory angle of the SST configuration $\theta=50.5^{o}$
$E_{lab}^{di-n}=3.12$~MeV and its energy detected by the TOF system will
be  $1.56$~MeV.
In Fig.~\ref{fig11},  positions of breakup events
(S-curve) in the plane of kinetic energies $E_1-E_2$ of two outgoing neutrons
detected in coincidence  are shown together with
positions of di-neutron energies determined  by the detection system
for SST configurations at $E=13, 19,$ and $65$~MeV.
Also the range of S-curve covered by data is indicated by two circles. Since
 at $13$~MeV di-neutrons  would come nearest to the SST region,
the data at that energy would be influenced  most by the
assumptive background of accidental coincidences.   
 The intensity of accidental coincidences depends on the number of
neutrons or di-neutrons arriving at detectors, which 
is determined by the energy spectra of outgoing neutrons in
incomplete $d(n,n)np$ breakup and by 
laboratory  angular distribution of di-neutrons from the $d(n,di-neutron)p$
reaction. The predictions for these quantities based on solutions of
the 3N Faddeev equation 
with a modified (in the $^1S_0$ partial wave)
 CD~Bonn potential   \cite{dineu_3ncont} 
at the above considered three  energies are shown in Fig.~\ref{fig12}. Since the
 cross section for di-neutrons production
is comparable to the cross section for production of neutrons in incomplete
nd breakup, it indeed seems plausible that accidental coincidences could
impact the measured  low-energy SST cross sections.
 Rapid diminishing of di-neutrons production with energy (see Fig.~\ref{fig12})
would also explain why the discrepancy between SST data
and theory decreases with energy. 
  New measurements of the low-energy nd SST cross sections as well as
  a measurement of nd SST at $E=65$~MeV,  
  using a detection system able to distinguish between neutrons and
  hypothetical di-neutrons, like the one
  proposed in \cite{bodek_2013}, would be very welcome.
 
\section{Summary and Conclusions}
\label{summary}

In this investigation we performed a comprehensive analysis of the available
SST Nd breakup cross section data using high precision (semi)phenomenological
NN potentials alone or combined with the standard 3N forces as well
 as selected  
 chiral forces. Four different chiral NN potentials including the
most precise SMS N$^4$LO$^+$ of Ref.~\cite{preinert} have been applied
alone or in combination with chiral 3NF's at different orders of chiral
expansion. The main results are summarized as follows.

\begin{itemize}

\item[-] The available nd SST data cover the range of incoming neutron
  lab. energies 
  between $E=10-25$~MeV while the pd data were measured for proton
  energies in a region  $E=10-65$~MeV. 
  The experiments  were performed by different groups using
  different experimental arrangements or techniques.
  When at a particular energy  several pd
  or nd data sets are available, the data from different
  measurements are consistent with each other. 

\item[-] Using (semi)phenomenological NN potentials alone or accompanied
  by the TM99 or Urbana IX 3NF one is not able to explain the low-energy SST
  pd and nd SST data. All theoretical predictions
  practically overlap in nd and pd systems with 
  pd data  overestimated by $\approx 10$~\%  and nd data underestimated
  by $\approx 20 - 30$~\%. The discrepancies between theory and 
  data diminish with
  increasing energy of the incoming nucleon. At $E=65$~MeV 
  NN force  predictions agree with the pd SST cross sections, while 
  inclusion of 3NF provides a slight overestimation of the data.

\item[-] Predicted low-energy SST cross sections based on
  different chiral NN potentials are independent from
  the type of regularization used or from the regularization parameters,
  and are practically
  identical to predictions of (semi)phenomenological interactions.
  At $E=65$~MeV that independence starts to be lost. 

\item[-] Adding the considered chiral  3NF at
  different orders of chiral expansion
   has no significant
  influence on the SST cross sections at low energy.
  Even a broader range of strengths from the correlation line ($c_D,c_E$)
  for N$^2$LO 3NF contact terms
     yields practically the same low-energy SST cross sections.
     Again, at $E=65$~MeV, this stability with respect
      to changes of strengths  vanishes and 3NF effects come into play. 

\item[-] The low-energy  SST cross sections originate practically from
  $^1S_0$ and $^3S_1-^3D_1$ NN force components. Changes introduced by
  a simple multiplication of the corresponding matrix elements by a factor
  $\lambda$ could explain the SST pd data but not nd ones. However, the
  required changes of the $^1S_0$ and/or $^3S_1-^3D_1$ are excluded
  by the NN data and/or by the $^3$H binding energy.

\item[-] In view of the astonishing stability of the low-energy SST cross
  sections to the underlying dynamics it seems very unlikely that
  charge symmetry breaking mechanism of any conceivable
  kind in 2N or 3N forces
  could be able to explain the low-energy nd SST cross sections.
  The explanation of the nd data should be thus sought in some exotic
  phenomena such as
  e.g. the hypothetical bound state of two neutrons.
  Supposable existence of the di-neutron would provide
  additional background in the region of the SST breakup phase-space, which
  could not be discerned in measurements performed so far.

\end{itemize}
  
  Further investigations and theoretical as well as experimental efforts
  are required to solve that low-energy SST puzzle. 
  From the experimental side, measurements of nd SST cross sections at
  low-energies with experimental arrangement able to 
  discern supposable di-neutron background would be needed.
  Also dedicated pd SST measurement directed to determine precise
  normalization of SST cross sections would be welcome. 
  From the theoretical side, efforts to fully include in 3N continuum
  calculations consistently regularized
  N$^3$LO and N$^4$LO 3NF components are required. This is the aim of
  the LENPIC project.

\begin{acknowledgments}
This study has been performed within Low Energy Nuclear Physics
International Collaboration (LENPIC) project and 
was  supported by the Polish National Science Center 
under Grant No. 2016/22/M/ST2/00173, by Deutsche Forschungsgemeinschaft (DFG)
and Natural Science Foundation of China (NSFC) through funds provided to the
Sino-German CRC 110 ``Symmetries and the Emergence of Structures in QCD''
(NSFC Grant No. 11621131001, DFG Project-Id 196253076-TRR 110) and by
Bundesministerium f\"ur Bildung und Forschung (BMBF), Grant No. 05P18PCFP1.  
 The numerical calculations were performed on the 
 supercomputer cluster of the JSC, J\"ulich, Germany.
 We would like to thank other members of the LENPIC Collaboration for
 interesting discussions and A. Nogga and K. Hebeler for providing us
 with matrix elements of N$^2$LO chiral 3NF's. 
\end{acknowledgments}


\bibliography{apssamp}

\newpage
\newpage
\newpage
\newpage
\newpage
\newpage
\newpage
\newpage
\newpage
\newpage
\newpage
\newpage

\end{document}